\documentclass[prb,twocolumn,aps,amsmath,amssymb]{revtex4}
\usepackage[T1]{fontenc}
\usepackage[utf8]{inputenc}
\usepackage{graphicx}
\usepackage{amsmath}
\usepackage{amssymb}
\usepackage{dcolumn}
\usepackage{float}
\usepackage{bm}
\usepackage{amsfonts,times}
\usepackage{xcolor}
\usepackage[breaklinks=true,colorlinks,citecolor=blue,linkcolor=blue,urlcolor=blue]{hyperref}

\DeclareMathAlphabet{\bi}{OML}{cmm}{b}{it}

\def\be{\begin{equation}}
\def\ee{\end{equation}}
\def\bearr{\begin{eqnarray}}
\def\eearr{\end{eqnarray}}

\begin{document}

\title{Conductance properties of $\alpha$-$T_3$ Corbino disks}
\author{Mijanur Islam}
\email[]{mislam@iitg.ac.in}

\author{Saurabh Basu}
\email[]{saurabh@iitg.ac.in}

\affiliation{Department of Physics, Indian Institute of Technology-Guwahati, Guwahati-781039, India.}

\normalsize
%\date{\today}

\begin{abstract}
In this work, we investigate an $\alpha$-$T_3$ lattice in the form of a Corbino disk, characterized by inner and outer radii $R_1$ and  $R_2$, threaded by a tunable magnetic flux. Through exact (analytic) solution of the stationary Dirac-Weyl equation, we compute the transmission probability of the carriers and hence obtain the conductance features for $0<\alpha \leq 1$ ($\alpha$ denotes the strength of the hopping between the central atom and one of the other two) which allows ascertaining the role of the flat band, alongwith scrutinizing the transport features from graphene to a dice lattice. Our results reveal periodic Aharonov-Bohm (AB) oscillations in the conductance, reminiscent of the utility of the Corbino disk as an electron pump. Further, these results are strongly influenced by parameters, such as, doping level, ratio of the inner and outer radii, magnetic flux, and $\alpha$. Additionally, complex quantum interference effect resulting in the possible emergence of higher harmonic modes and split-peak structures in the conductance, become prominent for smaller $\alpha$ values and larger ratios of the radii. We also find that, away from the charge-neutrality point (zero doping), the conductance oscillations are more pronounced and sensitive to the various parameters, with the corresponding behaviour largely governed via the evanescent wave transport. Further, the Fano factor reveals distinct transport regimes, transitioning from Poissonian to pseudo-diffusive for $\alpha < 1$, and from ballistic to pseudo-diffusive at the dice limit ($\alpha = 1$). Thus, this setup serves as a fertile ground for studying the generation of quantum Hall current and Aharonov-Bohm (AB) oscillations in a flat band system, alongwith demonstrating intricate appearance of higher harmonics in the electron transport. Finally, to put things in perspective, we have compared our results with those for graphene disks that highlight the difference between the two with regard to device applications.
 
\end{abstract}

\maketitle

\section{Introduction}
Two-dimensional (2D) Dirac materials owing to their massless fermionic dispersion and high electronic mobility have prospects providing the building block for future electronic devices. The striking electronic properties can be modified by nanostructuring and patterning, such as, manufacturing nanoribbons [\onlinecite{Par}], nanorings [\onlinecite{Rus}], junctions [\onlinecite{Wil}], quantum dots (QDs) [\onlinecite{Pon}], QD arrays [\onlinecite{Vai,Car}]. Therefore, the transport properties heavily depend on the design of the geometry, its edge shape, [\onlinecite{Gun,Zha}] etc. Here, we present one such annular geometry called as the Corbino disk which is contemplated by Laughlin as an electron pump that shows generation of radial (Hall) current as the enclosed flux quantum changes.

Recent studies have extensively investigated various quantum transport phenomena in graphene Corbino disks, both experimentally [\onlinecite{Yan,Pet,Kum,Zen,Liu}] and theoretically [\onlinecite{Ryc2,Ryc3,Kat,Kha,Abd,Jon,Sus,Ryc}]. The edge-free geometry of the disk enables transport studies via evanescent waves in nanoscale graphene systems [\onlinecite{Kat1}]. At zero magnetic field, the conductance of ultraclean ballistic disks as a function of the carrier concentration [\onlinecite{Kum}] aligns well with the basic mode-matching analysis described in Ref. [\onlinecite{Ryc2}]. Further, in presence of a non-zero magnetic field, periodic (approximately sinusoidal) magnetoconductance oscillations have been predicted [\onlinecite{Ryc3,Kat}], yet experimental confirmation of this intriguing quantum interference phenomenon is still lacking. Recently, the authors of Ref. [\onlinecite{Ryc}] demonstrated that a Corbino graphene disk, when pierced by a current carrying solenoid, can exhibit Aharonov-Bohm (AB) type conductance oscillations. They also found that these oscillations are more pronounced in the presence of an electrostatic potential, which breaks cylindrical symmetry and introduces mode mixing. Furthermore, Refs. [\onlinecite{Bou1,Bou}] explored the effects of mass and wedge disclination on a graphene Corbino disk, revealing that the {\it mass} term can conspire to eliminate tunneling effects by creating points of zero transmission. 

Soon after the emergence of graphene research, the Dirac-cone physics was combined with flat-band systems in a modified lattice known as the $\alpha$-$T_3$ lattice. This lattice is formed by coupling one of the inequivalent sites of the honeycomb lattice to an additional atom located at the center of the hexagon, with a coupling strength $\alpha$ [\onlinecite{Sut,Vid,Dor}]. This design effectively interpolates between graphene ($\alpha=0$) and the dice lattice ($\alpha=1$). Notably, the flat band intersects the nodal Dirac points, leading to unique phenomena such as an $\alpha$-dependent Berry phase [\onlinecite{Rao}], super-Klein tunneling [\onlinecite{She,Urb}], Weiss oscillations [\onlinecite{Fir}], etc. Additionally, the presence of flat bands enhances the magneto-optical and magnetotransport responses [\onlinecite{Che,Bis}]. In particular, analyses of the frequency-dependent magneto-optical and zero-field conductivity of Hg$_{1-x}$Cd$_x$Te [\onlinecite{Orl}] at the critical Cd concentration ($x_c \simeq 0.17$) has demonstrated its close association with the $\alpha$-$T_3$ model with $\alpha = 1/\sqrt{3}$ [\onlinecite{Mal}]. Other realizations of the $\alpha$-$T_3$ and dice systems can be fabricated in heterostructures of cubic lattices, such as SrTiO$_3$/SrIrO$_3$/SrTiO$_3$, as well as in cold bosonic or fermionic atoms trapped in optical lattices [\onlinecite{Riz,Wan}] etc.

In recent years, a number of properties of lower-dimensional systems in the $\alpha$-$T_3$ lattice have been investigated. These studies include the size dependence of energy levels in quantum rings (QR) [\onlinecite{Isl}], persistent currents in $\alpha$-$T_3$ QRs [\onlinecite{Isl1,Isl2}], the distribution of edge currents in QDs [\onlinecite{Son}], the emergence of Majorana corner states in the presence of Rashba coupling [\onlinecite{Moh}], an analysis of atomic effects in narrow zigzag ribbons [\onlinecite{Hao}], valley filtering [\onlinecite{Fil}], and the dynamic formation of bound states under an external magnetic field [\onlinecite{Fil1,Fil2}] in $\alpha$-$T_3$ QDs, as well as size effects in atomic collapsed dice lattices [\onlinecite{Ori}]. However, an $\alpha$-$T_3$ Corbino disk and its transport properties are yet to be explored. Thus, studying the electronic properties of the $\alpha$-$T_3$ Corbino disk presents an intriguing opportunity of examining the transmission coefficients and hence electrical conductance in presence of a threaded magnetic field or a solenoid. We will follow the Corbino disk setup proposed for graphene in Ref. [\onlinecite{Ryc}]. Following a general introduction of the model, we have outlined a solution with a view to obtain the transport features of the $\alpha$-$T_3$ Corbino disk.

\begin{figure}[h!]
\centering
\includegraphics[width=1.05\linewidth, height=0.7\linewidth]{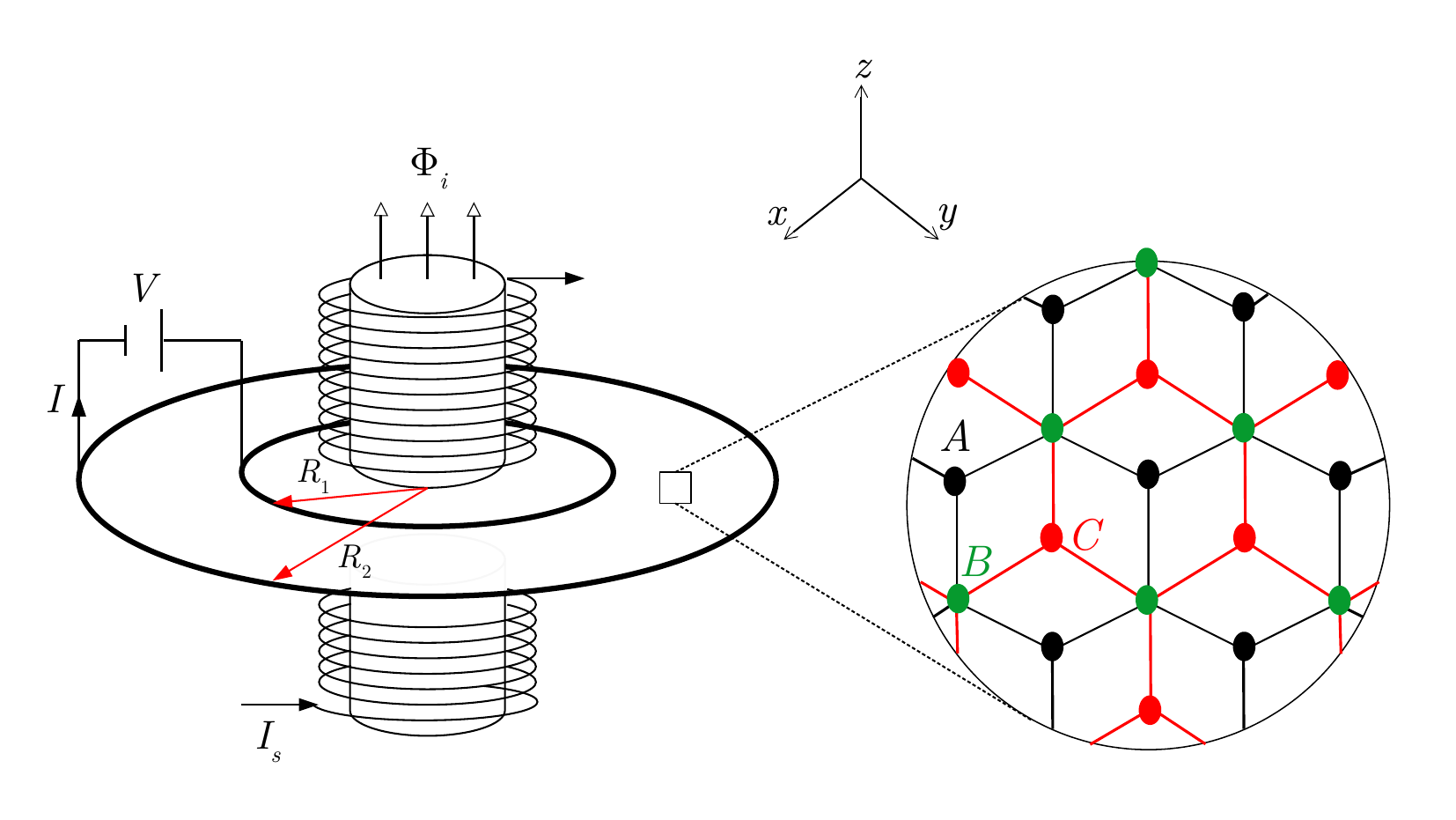}
\caption{(Color online) Schematic of an $\alpha$-$T_3$ Corbino disk with inner radius $R_1$ and outer radius $R_2$, connected by two electrodes as shown by thick black circles. A voltage source ($V$) drives the current ($I$) through the disk. A long solenoid, carrying the current $I_s$, generates the flux $\Phi_i$ piercing the annular regime. The coordinate system is also shown. The structure of the $\alpha$-$T_3$ lattice is depicted in the zoomed portion. $A$, $B$, and $C$ sub lattice sites are shown by black, green, and red dots, respectively. The hopping between the $A$ and $B$ sublattice is $t$ and between the $B$ and $C$ is $\alpha t$.}
\label{fig:sys}
\end{figure}

The manuscript is organized as follows. In Sec. \ref{For}, we introduce the $\alpha$-$T_3$ model, discuss the method of mode-matching analysis for the system as shown in Fig. \ref{fig:sys} and solve the eigenvalue problem for the $\alpha$-$T_3$ Corbino disk. Next, in Sec. \ref{Res}, the numerical computations of the transmission, conductance oscillations, and Fano factor are presented in Sec. \ref{Fano1}. A comparison between graphene and $\alpha$-$T_3$ Corbino disk is discussed in Sec. \ref{Compare} The conclusions of the main results are given in Sec. \ref{Con}.  

\section{Formalism}
\label{For}
Let us begin by considering a Corbino disk with an $\alpha$-$T_3$ lattice geometry, and characterized by an inner radius $R_1$ and outer radius $R_2$ which are connected by two electrodes. A current-carrying long solenoid (to eliminate edge effects) penetrates the inner area of the disk, yielding a magnetic flux $\Phi_i$ threading the annular region.

\begin{figure}[h!]
\centering
\includegraphics[width=1.05\linewidth]{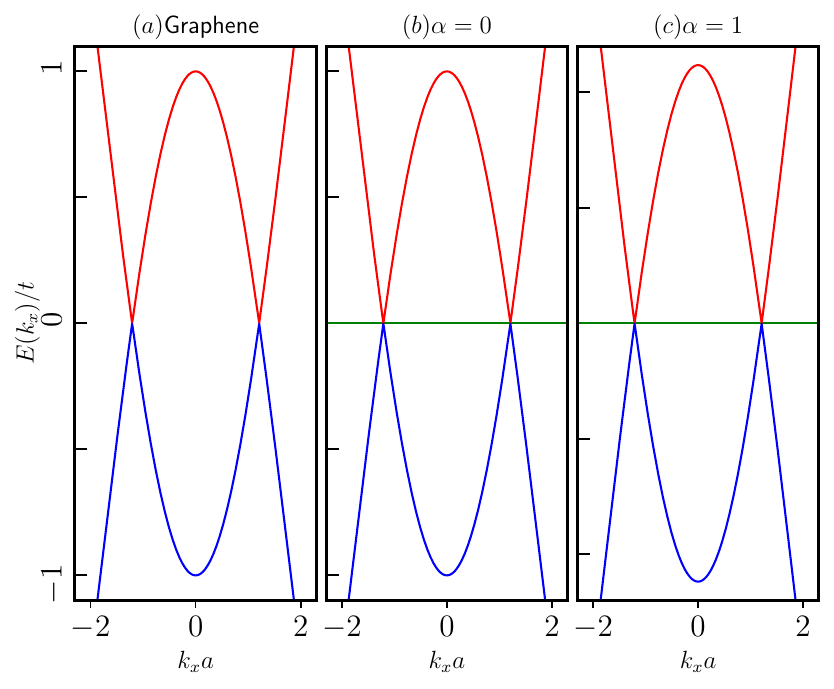}
\caption{(Color online) Bulk band structure of the $(a)$ graphene (two sublattice model), and $\alpha$-$T_3$ lattice with $(b)$ $\alpha=0$ (in three sublttice model), and $(c)$ $\alpha=1$ with $k_ya=2\pi/3$. For $\alpha$-$T_3$ lattice, the band structure is independent of $\alpha$, however they are normalised with a factor 1/$\cos\varphi$. There exists a zero energy flat band even if $\alpha=0$. The band structure is independent of $\alpha$. The band structure consists of
dispersive conduction and valence bands and a non-dispersive zero energy flat band. The conduction band touches the valence band at the Dirac points in the first Brillouin zone. Around those points, the spectrum becomes linearly dispersive. Further, the zero energy flat band exists even if $\alpha=0$.}
\label{fig:band}
\end{figure}

The bulk band structure of the $\alpha$-$T_3$ lattice is shown in Fig. \ref{fig:band}, which consists of dispersive conduction and valence bands and a non-dispersive zero energy flat band. The conduction band touches the valence band at the Dirac points in the first Brillouin zone. Further, in the vicinity of a specific valley (say, the $K$-valley), the low energy excitations in the $\alpha$-$T_3$ lattice can be described by the following Dirac-Weyl Hamiltonian
\begin{eqnarray}\label{Ham1}
H=v_F(\pi_xS_x+\pi_yS_y) + U(r),
\end{eqnarray}
where $v_F$ is the Fermi velocity and $\pi_x$ and $\pi_y$ are the components of the canonical momentum operator defined via ${\bm \pi}={\bm p}+e{\bm A}$. Here, $\bm p$ denotes the in-plane mechanical momentum operator ($-i\hbar (\partial_x, \partial_y)$) and the vector potential $\bm{A}$. Here, the components of the pseudospin-1 operator $\bm S$ associated with the $\alpha$-$T_3$ lattice 
are given by
\begin{equation*}
\resizebox{\hsize}{!}{$
S_x = \begin{pmatrix}
0 & \cos\phi & 0\\
\cos\phi & 0 & \sin\phi\\
0 & \sin\phi & 0
\end{pmatrix},
S_y =\begin{pmatrix}
0 & -i\cos\phi & 0\\
i\cos\phi & 0 & -i\sin\phi\\
0 & i\sin\phi & 0
\end{pmatrix}$}
\end{equation*}
with $\tan\phi=\alpha$. The $z$-component of ${\bm S}$ can be directly derived from the commutation relation $[S_x, S_y] = iS_z$. Due to the circular symmetry, we adopt the symmetric gauge for the vector potential, expressed as
\begin{eqnarray}
{\bm A} = (A_x, A_y) = \frac{\Phi_i}{2\pi}\left(-\frac{y}{r^2}, \frac{x}{r^2} \right).
\end{eqnarray}
We consider the case of a disk with an annular region pierced by a long solenoid, generating a flux $\Phi_i$. Additionally, we assume that the electrostatic potential energy $U(r)$ depends only on the radial distance, $r = \sqrt{x^2 + y^2}$. Specifically, $U(r) = 0$ within the disk region $(R_1 < r < R_2)$ and $U(r) = U_\infty$ (will be set to infinity) outside this area. The Hamiltonian in equation (\ref{Ham1}) commutes with the $z$-component of the total angular momentum operator, defined as $J_z = L_z + S_z$, where $L_z = -i\hbar \frac{\partial}{\partial \theta}$ is the orbital angular momentum operator, and $S_z$ is the pseudospin operator. The energy eigenfunctions can therefore be chosen as eigenstates of $J_z$, as
\begin{eqnarray}\label{EigT3}
\psi_j(r,\theta)=\begin{pmatrix}
\chi_{j,A}(r)e^{i(j-1)\theta}\\
\chi_{j,B}(r)e^{ij\theta}\\
\chi_{j,C}(r)e^{i(j+1)\theta}
\end{pmatrix},
\end{eqnarray}
where $j$ is an integer, the three spinor components ($A, B, C$) correspond to the sublattice degrees of freedom, with ($r, \theta$) denotes the coordinates. The Dirac equation now can be written as $H_j(r)\chi_j(r) = E\chi_j(r)$, where $\chi_j(r) = [\chi_{j,A}(r), \chi_{j,B}(r), \chi_{j,C}(r)]^T$, and 
\begin{equation}\label{Ham2}
\begin{split}
H_j(r) = & -i\hbar v_FS_x\partial_r + U(r) \\
& + \hbar v_FS_y \begin{pmatrix}
\frac{j-1}{r}+\frac{e\Phi_i}{hr} & 0 & 0 \\
0 & \frac{j}{r}+\frac{e\Phi_i}{hr} & 0 \\
0 & 0 & \frac{j+1}{r}+\frac{e\Phi_i}{hr}
\end{pmatrix}.
\end{split}
\end{equation}
For a piecewise-constant potential energy $U(r)$ and in presence of a bias voltage, which implies a varying electron concentration (called as doping later) such that the electron energies satisfy $E > U(r)$, the eigenfunctions of $H_j(r)$ (equation \ref{Ham2}) for the incoming waves (propagating from $r = 0$) and outgoing waves (propagating from $r = \infty$) are, up to a normalization constant, given by 
\begin{eqnarray}\label{Eig2}
\chi_j^{inc} = \begin{pmatrix}
-i\cos\phi\chi_1(kr)\\
H_{\nu(j)}^{(2)}(kr)\\
-i\sin\phi\chi_2(kr)
\end{pmatrix},
\chi_j^{out} = \begin{pmatrix}
-i\cos\phi\chi_3(kr)\\
H_{\nu(j)}^{(1)}(kr)\\
-i\sin\phi\chi_4(kr)
\end{pmatrix}
\end{eqnarray}
where
\begin{eqnarray}\label{nu}
\nu(j) = j+\Phi_i/\Phi_0,
\end{eqnarray}
\begin{eqnarray*}
\begin{split}
\chi_{1,2}(kr) &=\frac{H_{\nu(j)-1}^{(2)}(kr) - H_{\nu(j)+1}^{(2)}(kr)}{2} \pm \frac{\nu(j)H_{\nu(j)}^{(2)}(kr)}{kr},\\
&= \pm H_{\nu(j)\mp 1}^{(2)}(kr)
\end{split}
\end{eqnarray*}
\begin{eqnarray}
\begin{split}
\chi_{3,4}(kr) &=\frac{H_{\nu(j)-1}^{(1)}(kr) - H_{\nu(j)+1}^{(1)}(kr)}{2} \pm \frac{\nu(j)H_{\nu(j)}^{(1)}(kr)}{kr},\\
&= \pm H_{\nu(j)\mp 1}^{(1)}(kr)
\end{split}
\end{eqnarray}
$H_{\nu}^{(1,2)}(\rho)$ is the Henkel function of the first and second kinds, $\Phi_0$ is the usual flux quantum and $k = |E - U(r)|/(\hbar v_F)$. The + sign applies to  $\chi_1$ and $\chi_3$ and - sign applies to  $\chi_2$ and $\chi_4$ respectively. We use the relation $H_{\nu +1}^{(1,2)}(\rho) + H_{\nu -1}^{(1,2)}(\rho)= \frac{2\nu}{\rho}H_{\nu}(\rho)$. The solution for the Eq. (\ref{Ham2}) pertaining to the disk area can be represented as
\begin{eqnarray}
\chi_j^{(d)} = A_j \chi_j^{inc}(k_Fr) + B_j\chi_j^{out}(k_Fr), \;\;\;\;\;\; R_1<r<R_2
\end{eqnarray}
with  $A_j$ and $B_j$ being arbitrary constants, and the Fermi wave vector is given by $k_F = |E|/(\hbar v_F)$. 

The heavily doped (large bias voltage) $\alpha$-$T_3$ leads are modeled via taking the limit of $U(r) = U_\infty \to \pm \infty$ for $r < R_1$ or $r>R_2$. The corresponding wave functions in different regimes can be expressed as
\begin{eqnarray}
\chi_j^{(1)} = \frac{e^{\pm ik_\infty r}}{\sqrt{r}}\begin{pmatrix}
\cos\phi \\
1 \\
\sin\phi
\end{pmatrix}
+ r_j\frac{e^{\mp ik_\infty r}}{\sqrt{r}}\begin{pmatrix}
\cos\phi \\
-1 \\
\sin\phi
\end{pmatrix}, \;\; r<R_1
\end{eqnarray}

\begin{eqnarray}
\chi_j^{(2)} = t_j\frac{e^{\pm ik_\infty r}}{\sqrt{r}}\begin{pmatrix}
\cos\phi \\
1 \\
\sin\phi
\end{pmatrix}, \;\;\;\; r>R_2
\end{eqnarray}
where we have introduced the reflection (transmission) amplitudes $r_j (t_j)$ and $k_\infty = |E - U_\infty|/(\hbar v_F)\to \infty$. In the first term of Eq. (9) of the manuscript, $+ k_\infty$ denotes an incoming conduction electron from the lead, while the $-k_\infty$ corresponds to an incoming valence electron from the lead. In the second term of Eq. (9), $- k_\infty$ represents a reflected conduction electron from the edge of the disk, whereas the $+ k_\infty$ corresponds to a reflected valence electron from the edge of the disk. Similarly, in Eq. (10) of the manuscript, $+ k_\infty$ and $- k_\infty$ denote the transmitted conduction and valence electrons, respectively, through the edge of the disk. To proceed with the computation, we recall some useful formulae of the Hankel functions, namely,
\begin{eqnarray}
H_{\nu+\tau}^{(1)}(kr) = \mp \tau i H_\nu^{(1)}(kr),\;\; H_{\nu+\tau}^{(2)}(kr) = \pm \tau i H_\nu^{(2)}(kr),
\end{eqnarray}
where $\tau = \pm 1$. The mode-matching conditions, $\chi_j^{(1)}(R_1) = \chi_j^{(d)} (R_1)$ and $\chi_j^{(d)}(R_2) = \chi_j^{(2)} (R_2)$, enforcing the continuity of the wave functions at $R_1$, for the $B$ sublattice [\onlinecite{Fil2}],
\begin{eqnarray}
\chi_{j,B}^{(1)}(R_1) = \chi_{j,B}^{(d)}(R_1)
\end{eqnarray}
and for the $A$ and $C$ sublattices,
\begin{eqnarray}
\resizebox{\hsize}{!}{$
\cos\phi\chi_{j,A}^{(1)}(R_1) + \sin\phi\chi_{j,C}^{(1)}(R_1) = \cos\phi\chi_{j,A}^{(d)}(R_1) + \sin\phi\chi_{j,C}^{(d)}(R_1)$}.
\end{eqnarray}
Similarly, at $R_2$, for $B$ sublattice,
\begin{eqnarray}
\chi_{j,B}^{(d)}(R_2) = \chi_{j,B}^{(2)}(R_2)
\end{eqnarray}
and for the $A$ and $C$ sublattices,
\begin{eqnarray}
\resizebox{\hsize}{!}{$
\cos\phi\chi_{j,A}^{(d)}(R_1) + \sin\phi\chi_{j,C}^{(d)}(R_1) = \cos\phi\chi_{j,A}^{(2)}(R_1) + \sin\phi\chi_{j,C}^{(2)}(R_1)$}.
\end{eqnarray}
Hence, we find the transmission probability for the $j$th mode to be given by
\begin{eqnarray}\label{Trans}
\resizebox{\hsize}{!}{$
\begin{split}
T_{\nu(j)} = & |t_{\nu(j)}|^2\\
 = & \frac{4}{(kR_1)(kR_2)}\frac{\frac{4}{\pi^2}+[\mathfrak{D}_{\nu(j)}^{(1)}]^2}{[\mathfrak{D}_{\nu(j)}^{(2)}]^2 + [\mathfrak{D}_{\nu(j)}^{(3)}]^2 + [\mathfrak{D}_{\nu(j)}^{(4)}]^2 + [\mathfrak{D}_{\nu(j)}^{(5)}]^2},
\end{split}$}
\end{eqnarray}
where $\nu(j)$ is given by Eq. (\ref{nu}) and 
\begin{eqnarray*}
\mathfrak{D}_{\nu(j)}^{(1)} = Im[H_{\nu(j)}^{(2)}(kR_2)-H_{\nu(j)}^{(1)}(kR_2)]\cos2\phi,
\end{eqnarray*}
\begin{eqnarray*}
\resizebox{\hsize}{!}{$
\mathfrak{D}_{\nu(j)}^{(2)} = Im[H_{\nu(j)}^{(1)}(kR_2)H_{\nu(j)}^{(2)}(kR_1) + H_{\nu(j)+1}^{(1)}(kR_2)H_{\nu(j)+1}^{(2)}(kR_1)]$},
\end{eqnarray*}
\begin{eqnarray*}
\resizebox{\hsize}{!}{$
\mathfrak{D}_{\nu(j)}^{(3)} = Im[H_{\nu(j)+1}^{(1)}(kR_2)H_{\nu(j)}^{(2)}(kR_1) - H_{\nu(j)}^{(1)}(kR_2)H_{\nu(j)+1}^{(2)}(kR_1)]$},
\end{eqnarray*}
\begin{eqnarray*}
\resizebox{\hsize}{!}{$
\begin{split}
\mathfrak{D}_{\nu(j)}^{(4)} = & \big(1+\nu(j)\big) \bigg(\frac{1}{kR_2}-\frac{1}{kR_1}\bigg)\cos2\phi\\
& Im[H_{\nu(j)}^{(1)}(kR_1)H_{\nu(j)}^{(2)}(kR_2) - H_{\nu(j)}^{(1)}(kR_2)H_{\nu(j)}^{(2)}(kR_1)],
\end{split}$}
\end{eqnarray*}
\begin{eqnarray}
\label{Eq.17}
\mathfrak{D}_{\nu(j)}^{(5)} =  \frac{\nu(j)^2\cos^22\phi}{(kR_1)(kR_2)}\mathfrak{D}_{\nu(j)}^{(3)}.
\end{eqnarray}
The lattice structure of the $\alpha$-$T_3$ lattice consists of three distinct sublattices, namely, $A$, $B$, and $C$. These sublattices are connected by two different hopping amplitudes, $t$ (between the $A$ and $B$ sublattices) and $\alpha t$ (between the $B$ and $C$ sublattices) (see Fig. \ref{fig:sys}). During electron transmission, an electron can propagate between two $B$ sublattices via two distinct paths, one is characterized by hopping strength, $t$, while the other by $\alpha t$. Whereas, for the $A$ and $C$ sublattices, the available hopping paths are limited to $t$ for the $A$ sublattice and $\alpha t$ for the $C$ sublattice, respectively. This path difference induces a phase difference, giving rise to interference effects for any nonzero value of $\alpha$. This interference mechanism is a fundamental distinction from the graphene case. These path differences derive support from the expressions of $\mathfrak{D}_{\nu(j)}^{(1)}$, $\mathfrak{D}_{\nu(j)}^{(4)}$, and $\mathfrak{D}_{\nu(j)}^{(5)}$ in Eq. (\ref{Eq.17}). These terms depend on the value of $\alpha$, highlighting the role of hopping asymmetry in the system.

\section{Results and discussions}
\label{Res}
\begin{figure}[h!]
\centering
\includegraphics[width=\linewidth]{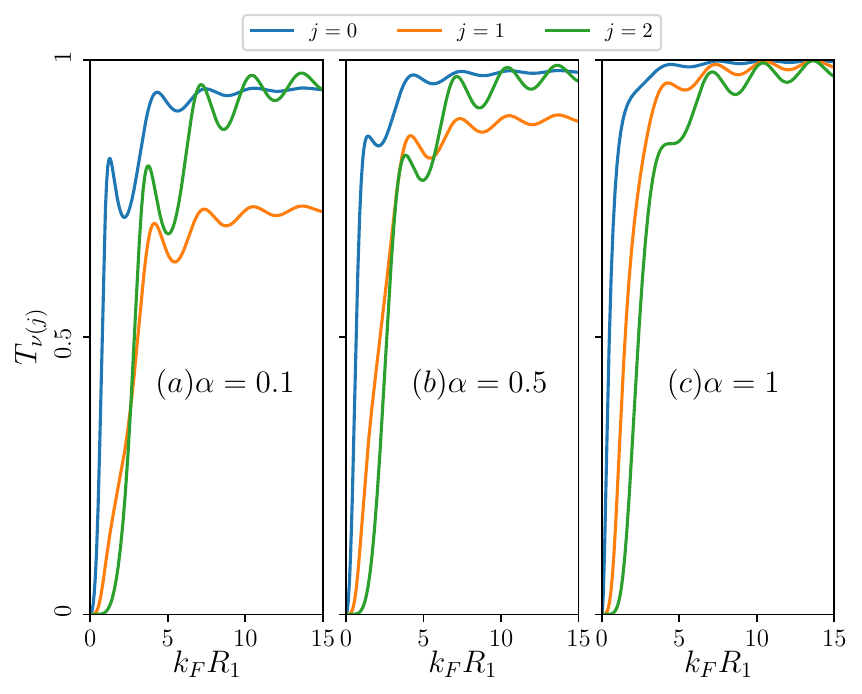}
\caption{(Color online) The transmission $T_{\nu(j)}$ ($j=0$, $1$, $2$) as a function of the doping parameter, $k_FR_1$ for a radii ratio $R_2/R_1=5$, $\Phi_i/\Phi_0=1/2$, for $(a)$ $\alpha=0.1$, $(b)$ $\alpha=0.5$, and $(c)$ $\alpha=1$.}
\label{fig:TvsR1}
\end{figure}
\begin{figure}[h!]
\centering
\includegraphics[width=\linewidth]{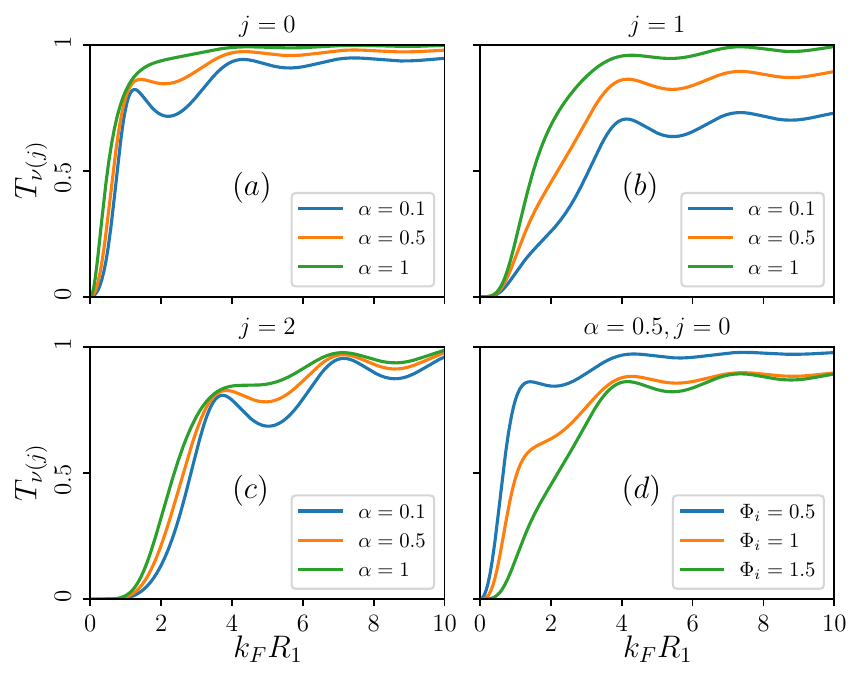}
\caption{(Color online) The transmission $T_{\nu(j)}$ as a function of the doping $k_FR_1$ for a radii ratio $R_2/R_1=5$, $\Phi_i/\Phi_0=1/2$, and for different values of $\alpha$ (different colors correspond to different values of $\alpha$) for $(a) j=0$, $(b) j=1$, and $(c) j=2$. $(d)$ Showing the transmission $T_{\nu(j)}$ as a function of the doping $k_FR_1$ for a radii ratio $R_2/R_1=5$, for a fixed value of the angular momentum quantum number, $j$ (namely, $j=0$), for a particular value of $\alpha$ (say, $\alpha=0.5$), and for different $\Phi_i/\Phi_0$ ratio as shown by different colors in the plot.}
\label{fig:TvsR1_m}
\end{figure}
\begin{figure}[h!]
\centering
\includegraphics[width=\linewidth]{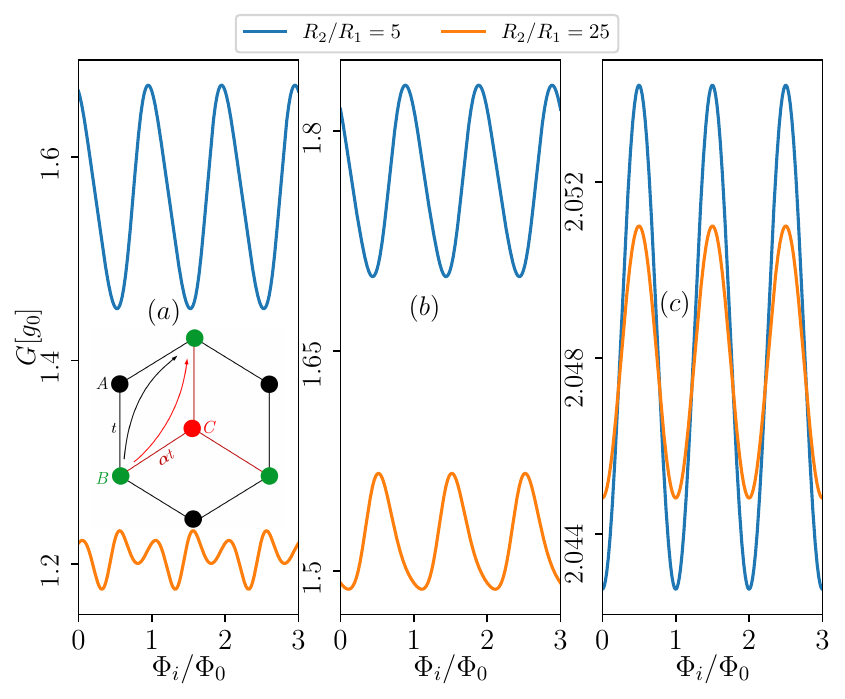}
\caption{(Color online) Conductance $G$ as a function of the flux piercing the annular region ($\Phi_i/\Phi_0$) for doping $k_FR_1=0.5$. The radii ratio taken as, $R_2/R_1=5$ (thin limit) and $25$ (thick limit), and specified for each lines on the plot with $(a)$ for $\alpha=0.1$, $(b)$ for $\alpha=0.5$, and $(c)$ for $\alpha=1$. In the inset of panel $(a)$, we illustrate the lattice structure along with the possible transmission paths for the $B$ sublattice, indicated by black and red arrows.}
\label{fig:Gvsphi_R2}
\end{figure}

We study the transmission $T_{\nu(j)} (j = 0,1,2)$, the conductance $G$, and the Fano factor $F$ numerically for the $\alpha$-$T_3$ Corbino disk with an annular region at the center ($R_1<r<R_2$) and threading a flux through it as shown in Fig. \ref{fig:sys}. We show our results as a function of a dimensionless quantity, $k_FR_1$, which essentially controls the electron doping (through $k_F$). Fig. \ref{fig:TvsR1} represents the transmission as a function of $k_FR_1$ for three different values of $\alpha$, namely, $\alpha=0.1$, $0.5$, and $1$ in Figs. $(a), (b)$, and $(c)$ respectively, for a particular width of the $\alpha$-$T_3$ disk denoted by $R_2/R_1=5$ and for a certain flux $\Phi_i/\Phi_0=1/2$. The transmission characteristics are shown for three different values of the angular momentum quantum number (corresponding to $J_z$) $j$, namely, $j=$ 0, 1, and 2 each represented by different color as shown in the figure (see Fig. \ref{fig:TvsR1}). We observe that the transmission strongly depends upon $j$, $\alpha$, and the value of the magnetic flux. The transmission curves show a sharp increase for small values of $k_FR_1$. In Fig. \ref{fig:TvsR1}$(a)$, for $\alpha=0.1$ (low values of $\alpha$) the curves display oscillatory behaviour before they converge toward a saturation. This indicates that a small non-zero value of $\alpha$ introduces additional effects, possibly interference or resonance phenomena. The transmission for $j=0$ (blue curve) increases more rapidly, while for $j=1$ and $j=2$, the oscillations are more gradual, and they converge slower than the $j=0$ case. In Fig. \ref{fig:TvsR1}$(b)$ for $\alpha=0.5$, the oscillatory  behaviour becomes slightly more prominent compared to $\alpha =0.1$ (Fig. \ref{fig:TvsR1}$(a)$). The curves for larger doping saturate, however the oscillations for higher $j$ values (especially $j=2$) are more discernible. This suggest that the effect of increasing $\alpha$ is to minimize these oscillations, while still maintaining the overall trend of increasing transmission with doping. In Fig. \ref{fig:TvsR1}$(c)$, for $\alpha=1$, the transmission smoothes compared to the lower values of $\alpha$. While some oscillations are still visible, particularly for $j=1$ and $j=2$, they are less pronounced. For $j=0$, the transmission reaches its saturation faster than for $j=1$ and $j=2$ curves. As $\alpha$ increases, the transmission for a particular $j$ value increases which is shown in Figs. \ref{fig:TvsR1_m} $(a)$, $(b)$, and $(c)$ for $j=0,$ $1$, and $2$ respectively for $\alpha=0.1$, $0.5$, and $1$. However, as $\alpha$ increases, the transmission curves become less oscillatory, particularly for higher $j$ values. The saturation of transmission occurs for large $k_FR_1$, indicating that large doping eventually results in a maximum transmission that does not significantly change with further increases in $k_FR_1$. The saturation occurs due to the fact that the energy levels nearly form a continuous spectrum at large doping. Further, we have a full transmission for larger $\alpha$ with $j=0$. Larger the magnetic flux, the transmission decreases as shown in Fig. \ref{fig:TvsR1_m}$(d)$ for a fixed $j$ and for particular $\alpha$ (namely, $\alpha = 0.5$).
\begin{figure}[h!]
\centering
\includegraphics[width=\linewidth]{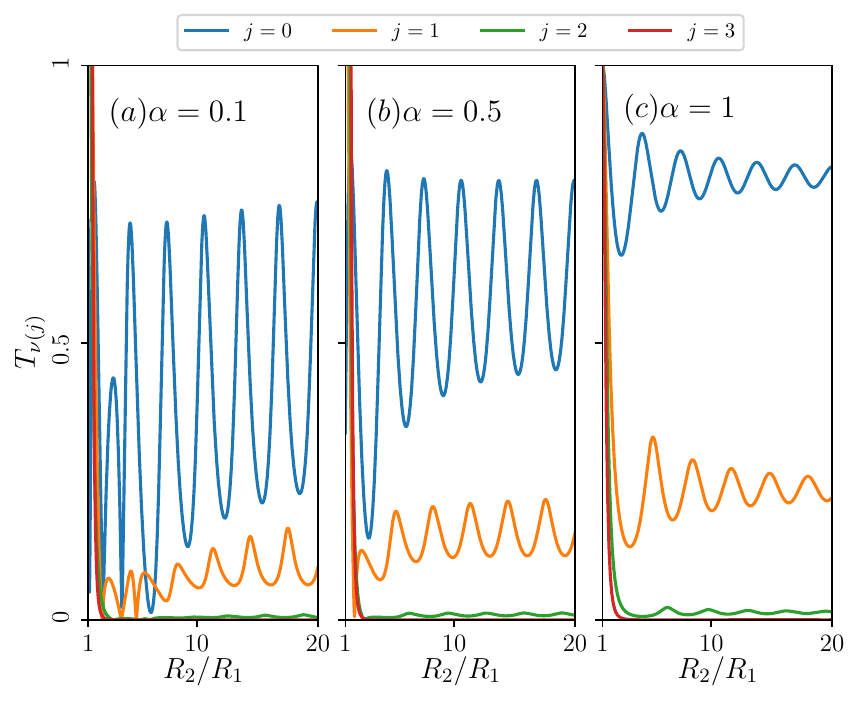}
\caption{(Color online) The transmission $T_{\nu(j)}$ ($j=0$, $1$, $2$, $3$) as a function of radii ratio $R_2/R_1$ for the doping $k_FR_1=1$, $\Phi_i/\Phi_0=1/2$, $(a)$ for $\alpha=0.1$, $(b)$ for $\alpha=0.5$, and $(c)$ for $\alpha=1$.}
\label{fig:TvsR2}
\end{figure}
Fig. \ref{fig:TvsR2} shows the transmission $T_{\nu(j)}$ as a function of the ratio of the outer and inner radii, namely, $R_2/R_1$ for a fixed doping value, namely, $k_FR_1=1$, with $\Phi_i/\Phi_0=1/2$ for $\alpha=0.1$, $0.5,$ and $1$. The transmissions are plotted for three different values of $j(=0,$ $1$, $2$, $3$). The transmission curves exhibit an oscillatory behaviour as $R_2/R_1$ increases. The curves corresponding to $j=0$ (blue lines) show largest transmission and are characterized by more prominent oscillations, with the transmission reaching close to 1 (complete transmission) at specific (lower) values of $R_2/R_1$. As $\alpha$ increases, the amplitudes and the frequency of these oscillation decrease.  Further, higher values of $j$ demonstrate the transmission to be significantly reduced and registers fewer oscillations. As $\alpha$ increases, higher $j$ curves exhibit more pronounced oscillations with an enhancement of the resonance effects. That can be understood from the lattice structure of the $\alpha$-$T_3$ lattice as shown in Fig. \ref{fig:sys}. The sublattices in an $\alpha$-$T_3$ lattice are connected with two different hopping amplitudes, namely $t$ (between the $A$ and $B$ sublattices) and $\alpha t$ (between the $B$ and $C$ sublattices). In a transmission process, an electron can travel between two $B$ sublattices through two distinct paths, one characterized by a hopping strength $t$ and another by $\alpha t$, as indicated by the black and red arrows in the inset of Fig. \ref{fig:Gvsphi_R2} ($a)$. Whereas, for the $A$ and $C$ sublattices, the available hopping paths are limited to $t$ for the $A$ sublattice and $\alpha t$ for the $C$ sublattice, respectively. This path difference introduces a phase difference, leading to interference effects for any nonzero value of $\alpha$. This interference mechanism is a key distinction from the graphene Corbino disk. As $\alpha$ increases, the path difference and corresponding phase difference decrease, leading to an enhancement in transmission. This effect highlights the tunability of electron transport in the $\alpha$-$T_3$ lattice. For $\alpha=1$, the transmission oscillations smoothen, particularly for $j=0$, where the transmission reaches a nearly constant value after some initial oscillations. For higher values of $j$ and all values of $\alpha$, the transmission decreases exponentially. We notice that the amplitudes of these oscillation decreases when the ratio $R_2/R_1$ is increased and further they depends on the value of $\alpha$.
\begin{figure}[h!]
\centering
\includegraphics[width=\linewidth]{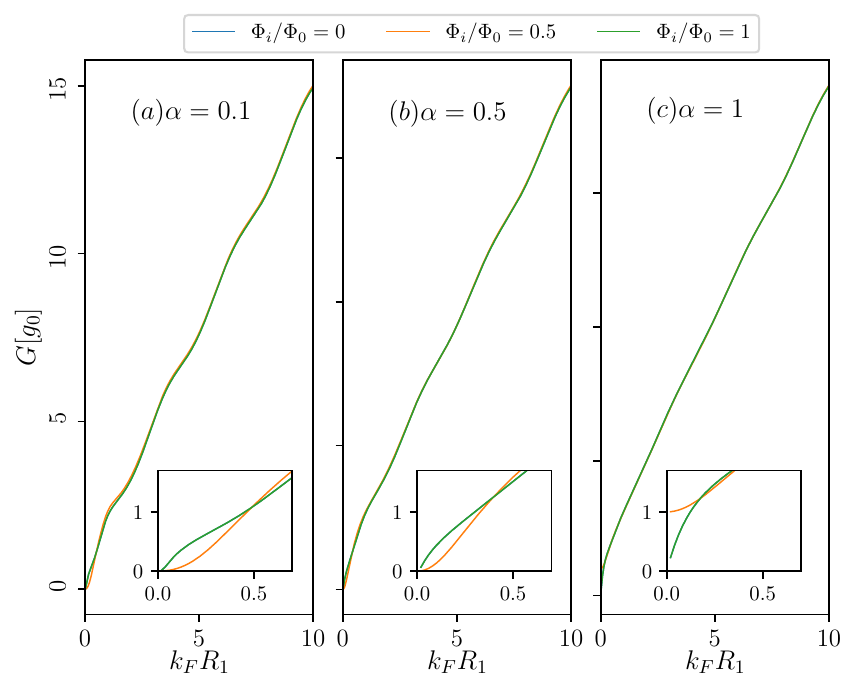}
\caption{(Color online) Conductance $G$ as a function of the doping $k_FR_1$ for the radii ratio $R_2/R_1 = 5$. Different lines corresponding to $\Phi_i=0$ (blue line), $\Phi_i=1/2$ (orange line), and $\Phi_i=1$ (green line) $\Phi_i$ is measured in unit of $\Phi_0$. Different plots correspond to $(a)$ $\alpha=0.1$, $(b)$ $\alpha=0.5$, and $(c)$ $\alpha=1$. In the insets, zoomed in views are shown.}
\label{fig:GvsR1}
\end{figure}

Now we compute certain physical quantities related to the transmission coefficient. The  conductance at the linear-response level is calculated via the Landauer-B$\ddot{\mathrm{u}}$ttiker formula [\onlinecite{Lan, But}]
\begin{eqnarray}\label{cond}
G = \frac{I}{V} = g_0 \sum_{j=0, \pm 1, \pm 2,...} T_{\nu(j)},
\end{eqnarray}
where the conductance quantum $g_0 = 4e^2/h$, with the factor 4 accounting for the spin and valley degeneracy, and the summation is performed over all the modes. In addition, the particle-hole symmetry, $T_{\nu(j)}(-k_FR_1)=T_{\nu(j)}(k_FR_1)$ allows us to limit the discussion only positive values of $k_FR_1$, that is, $k_FR_1>0$. Our numerical results are presented in the Fig. \ref{fig:GvsR1}.

The asymptotic properties of the Hankel functions [\onlinecite{Nem}] in Eq. (\ref{Trans}) lead to $T_{\nu(j)}\approx 1$ for $k_FR_1-\nu(j)\gg 1$, with $\nu(j)$ given by Eq. (\ref{nu}), or to $T_{\nu(j)}\approx 0$ for $\nu(j)-k_FR_1\gg 1$. With these, the conductivity can be approximated as $G\approx 2g_0k_FR_1$ for $k_FR_1\gg 1$ and $R_2\gg R_1$.

Let us put the ongoing discussion in perspective by mentioning some of the interesting results in graphene, and their relevance to those for the $\alpha$-$T_3$ Corbino disk. Previously, the conductance quantization with steps of $4e^2/h$ was predicted for a graphene strip with a moderate aspect ratio (width/length $\leq 1$) [\onlinecite{Bee, Yao, Tka, Ryc1}]. Quantization in unit of $8e^2/h$ was theoretically found in a bipolar junction in graphene, which exhibits the Goos-H$\ddot{\rm{a}}$nchen effect [\onlinecite{Bee1}]. The absence of conductance quantization observed in a graphene Corbino disk [\onlinecite{Ryc2, Ryc3}] highlights the role of the evanescent modes, which decay slowly with distance (following a power law) and remain significant far from the Dirac point, illustrating a striking consequence of angular momentum conservation [\onlinecite{Ryc2, Ryc3, Ryc, Bou, Bou1}]. A similar behaviour in the conductance profile is also observed here for the $\alpha$-$T_3$ Corbino disk as we see below.
\begin{figure}[h!]
\centering
\includegraphics[width=\linewidth]{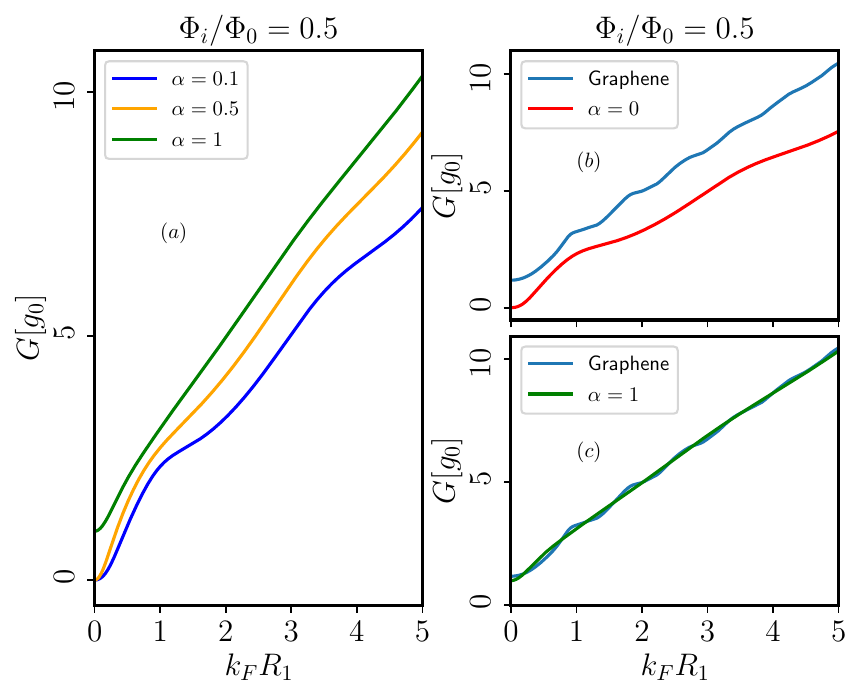}
\caption{(Color online) $(a)$ Conductance $G$ as a function of the doping $k_FR_1$ for the radii ratio $R_2/R_1 = 5$ and $\Phi_i/\Phi_0=0.5$ of the $\alpha$-$T_3$. Different lines corresponding to $\alpha=0.1$ (blue line), $\alpha=0.5$ (orange line), and $\alpha=1$ (green line). $(b)$ The same for the graphene (two sublattice model) and $\alpha=0$ case of the $\alpha$-$T_3$ lattice. $(c)$ The same for the graphene (two sublattice model) and $\alpha=1$ case of the $\alpha$-$T_3$ lattice.}
\label{fig:GvskF}
\end{figure}

Fig. \ref{fig:GvsR1} shows the conductance $G[g_0]$ as a function of doping ($k_FR_1$) for $\Phi_i/\Phi_0=0,$ $0.5$, and $1$ with the same $\alpha$ values as earlier, namely, 0.1, 0.5, and 1 in Figs. \ref{fig:GvsR1}$(a)$, $(b)$, and $(c)$ respectively. It is evident that the conductance $G$ increases with increase in doping, showing nearly step like behaviour as a function of $k_FR_1$. It may be noted that the parameter $k_FR_1$ is a dimensionless quantity that reflects the doping level in the system. As doping increases, a greater number of electronic states become available for transport, leading to an overall enhancement in the transmission characteristics, regardless of the values of $j$, $\alpha$, and $\Phi_i$. However, the sharpness of the steps diminishes as $\alpha$ increases. This suggests that conductance is (nearly) proportional to doping for higher values of $\alpha$. Additionally, for higher $k_FR_1$, all the three curves for $\Phi_i/\Phi_0 =0$, $0.5$, and 1 overlap significantly, showing very little dependence on the flux enclosed by the $\alpha$-$T_3$ disk. This also suggests that, except at very low doped regimes, the conductance becomes independent of the magnetic flux. In contrast, for very low $k_FR_1$ values, the differences between the curves for different $\Phi_i$ values are somewhat noticeable (see insets of Fig. \ref{fig:GvsR1}), with even those becoming somewhat unnoticeable for larger $\alpha$. Furthermore, the $\Phi_i/\Phi_0=0$ and $\Phi_i/\Phi_0=1$ curves are same, reminiscent of the AB effect in this disk-like geometry. Moreover, Fig. \ref{fig:GvskF}$(a)$ shows that increasing $\alpha$ leads to higher conductance at any doping level. When $\alpha=1$ (dice lattice), the conductance is most sensitive to the magnetic flux and demonstrates large conductance.
\begin{figure}[h!]
\centering
\includegraphics[width=\linewidth]{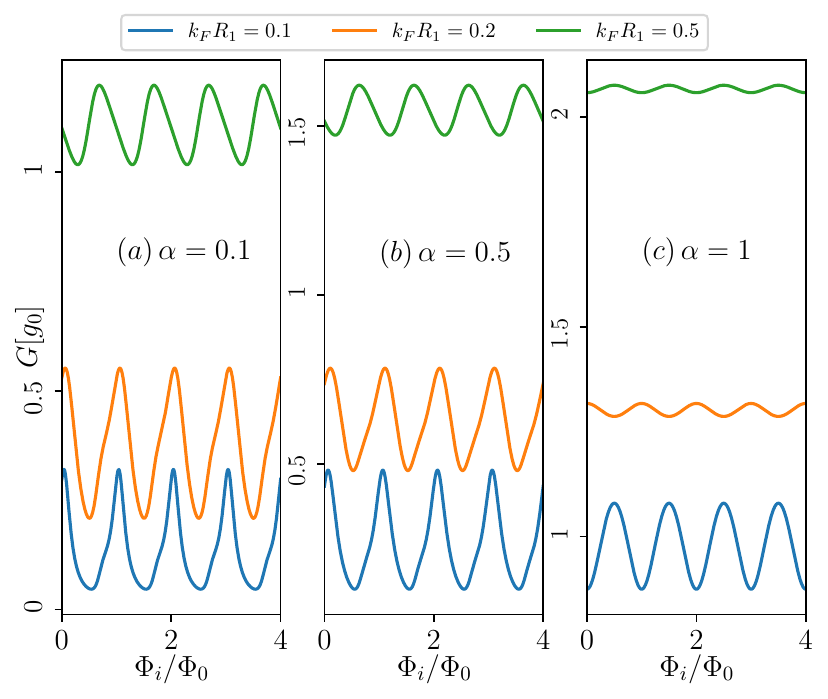}
\caption{(Color online) Conductance $G$ as a function of the flux piercing the annular region ($\Phi_i/\Phi_0$) for the radii ratio $R_2/R_1=5$. The doping varies from $k_FR_1=0.1$ to $0.5$ and specified for each lines on the plot. $(a)$ for $\alpha=0.1$, $(b)$ for $\alpha=0.5$, and $(c)$ for $\alpha=1$.}
\label{fig:Gvsphi}
\end{figure}

\begin{figure}[h!]
\centering
\includegraphics[width=\linewidth]{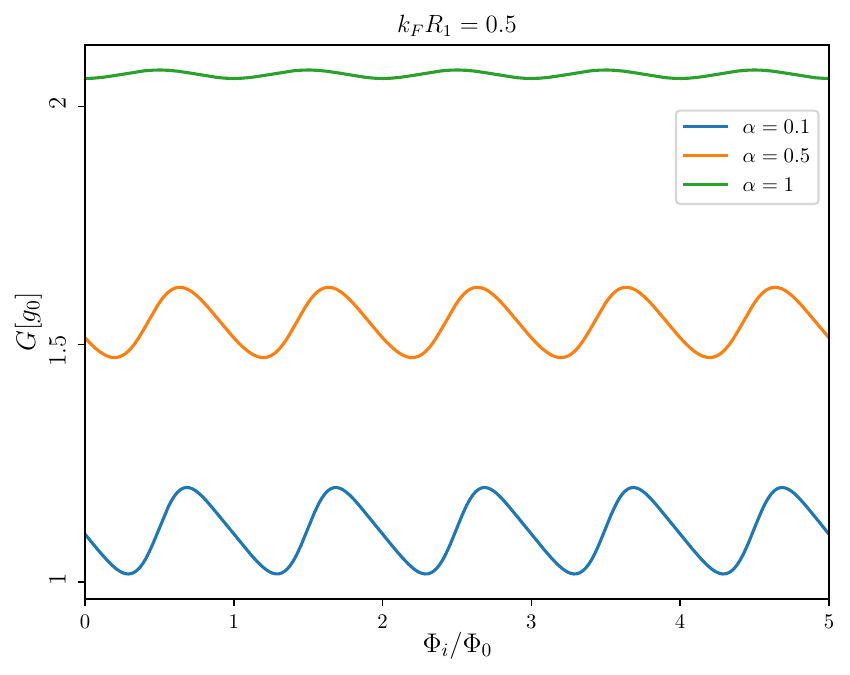}
\caption{(Color online) Conductance $G$ as a function of the flux piercing the annular region ($\Phi_i/\Phi_0$) for the radii ratio $R_2/R_1 = 5$ and $k_FR_1=0.5$. Different lines corresponding to $\alpha=0.1$ (blue line), $\alpha=0.5$ (orange line), and $\alpha=1$ (green line).}
\label{fig:Gvsphi1}
\end{figure}

The structure of the Eqs. (\ref{nu}), (\ref{Trans}), and (\ref{cond}) leads to a perfectly periodic functional dependence of $G(\Phi)$ at any arbitrary level of doping. In Figs. \ref{fig:Gvsphi}, \ref{fig:Gvsphi1}, and \ref{fig:Gvsphi_R2}, we show the conductance $G[g_0]$ as a function of the magnetic flux. It is clearly evident that the conductance oscillations are of the AB type with an oscillation period of one flux quantum, $\Phi_0$. The amplitudes of these oscillations depend on $k_FR_1$, $\alpha$, and the $R_2/R_1$ ratio. These results align well with the previously studied results for graphene [\onlinecite{Ryc,Bou,Bou1}]. The oscillations depict the quantum interference effects on the electron transport in mesoscopic disk structures, where the flux $\Phi_i$ through the disk modulates the conductance due to the AB effect. Here, the phase of the electronic wavefunction is affected by the presence of the magnetic flux (or the line integral of the vector potential), even if the electron travels through a region where the magnetic field is zero. The different panels illustrate the evolution of conductance oscillations with increasing $\alpha$. As $\alpha$ increases, the oscillations become less pronounced (a general trend seen throughout), and the amplitudes of the oscillations decrease for a particular value of $k_FR_1$ (here, $k_FR_1 =0.5$) as shown in Fig. \ref{fig:Gvsphi1}. This suggests that $\alpha$ influences certain aspects pertaining to the system geometry (or equivalently, the symmetry of the underlying $\alpha$-$T_3$ lattice), which in turn affects the quantum interference pattern. Further, increasing $\alpha$ softens the oscillations. The data points in each plot of Fig. \ref{fig:Gvsphi} represent different doping levels ($k_FR_1$). Doping affects the Fermi wavelength of the electrons, and consequently has implications on their interference pattern. Higher doping tends to reduce the prominence of the conductance modulation, as seen from the diminished oscillations corresponding to larger values of $k_FR_1$.

Fig. \ref{fig:Gvsphi_R2} illustrates the conductance $G[g_0]$ as a function of the flux $\Phi_i/\Phi_0$, with $k_FR_1=0.5$ (moderate doping), while varying the ratio of the radii $R_2/R_1$ (we have considered two limits, namely the {\it thin} limit ($R_2/R_1 = 5$) and the {\it thick} limit ($R_2/R_1=25$)), for different values of the hopping parameter $\alpha$. For a particular value of $\alpha$, as $R_2/R_1$ increases (as shown by the different curves in the plot), the amplitudes of the oscillations decrease. Additionally, for cases where $\alpha < 1$, there is a noticeable phase difference between different radius ratios. However, when $\alpha = 1$, the oscillations are in phase for both values $R_2/R_1$. Increasing the outer radius, $R_2$, expands the effective area of the disk region, which subsequently alters the interference patterns in the system.

The $\alpha=0$ (graphene) case, as discussed in the Refs. [\onlinecite{Ryc,Bou,Bou1}], demonstrates the conductance oscillations to be more regular, with a uniform amplitude and frequency with no asymmetry produced due to the absence of the central atom [\onlinecite{Bipar}], yielding ideal AB oscillations. With the inclusion of the central atom ($\alpha \neq 0$) especially for small values of $\alpha$ and large $R_2/R_1$ ratio, the oscillations cease to be smooth and become irregular (see Fig. \ref{fig:Gvsphi_R2}$(a)$). Instead, one can observe a double peak (or humps) in the conductance profile for each oscillation cycle, implying that there is an interference between different periodicities or frequencies.

This multi-harmonic contribution to the conductance oscillation may occur due to the availability of multiple pathways for the electrons contributing to the interference phenomena, such as the electron can traverse in both clockwise and anticlockwise paths around the flux as shown by the black and red arrows in the inset of Fig. \ref{fig:Gvsphi_R2}$(a)$. The hopping parameter $\alpha$ introduces a generic asymmetry in the system due to different coordination numbers for different sublattice sites [\onlinecite{Bipar}], while $\alpha=0$ represents a fully symmetric case. Even an infinitesimal value of $\alpha$ disturbs this symmetry, resulting in mixing of different modes leading to interference between them. As a result, we see a single peak turning complex and eventually evolving into a double-peak (split-peak) feature. Further, large $R_2/R_1$ implies larger area of the disk with larger number of sites and hence larger number of modes being involved. This may lead to different effective path lengths, resulting in a phase difference between them. This phase difference could result in a constructive interference at any two points within a single oscillation cycle, thereby producing the double-peak pattern.

Another possibility is that a larger outer radius (or larger $R_2/R_1)$ induces higher harmonic contributions to the conductance oscillations. These higher harmonics have shorter periods and may interfere with the fundamental AB oscillation periods to produce the observed double-hump structure. Further, the magnetic flux enclosed by different electron paths in the system could also vary due to large disk area, causing the effective flux experienced by electrons to differ depending on their paths. This difference can lead to superposition of more than one oscillation frequencies, producing two peaks within a single oscillation period. This behaviour is typical in mesoscopic systems where the geometry and the confinement phenomenon conspire to yield complex interference effects between electron paths.

As $\alpha$ increases, the double-peak feature in the conductance oscillations gradually disappears, even at {\it thick} limit. Further, as $\alpha$ increases, the system becomes progressively more asymmetric [\onlinecite{Wan}]. This larger asymmetry is likely to disrupt the coherent interference between different electron paths that produces the double-peak feature at lower values of $\alpha$. In the more asymmetric case (or a {\it thick} disk), the phase difference between the paths becomes more pronounced, and the interference effects that once contributed to the double-peak may cease to exist. Instead, the system possibly favours a more uniform interference pattern with regular AB-like oscillations. At higher $\alpha$, the interference from the higher harmonics diminishes or the additional modes become weaker such that they no longer constructively interfere with the primary oscillations. As a result, the primary modes (fundamental AB oscillations) begin to dominate, and the conductance oscillations tend to have more uniform periodicity. This simplifies the oscillation pattern and eliminates the double-peak structure. However, even at high $R_2/R_1$, the same feature disappears for all values of $\alpha$. This implies that the effect of increasing $\alpha$ is strong enough to suppress the complex (repeated) interference effects, regardless the size of the system. Further, the symmetry with respect to the direction of the flux has the consequence that the conductance is symmetric, that is, $G(-\Phi_i)=G(\Phi_i)$.

\begin{figure}[h!]
\centering
\includegraphics[width=\linewidth]{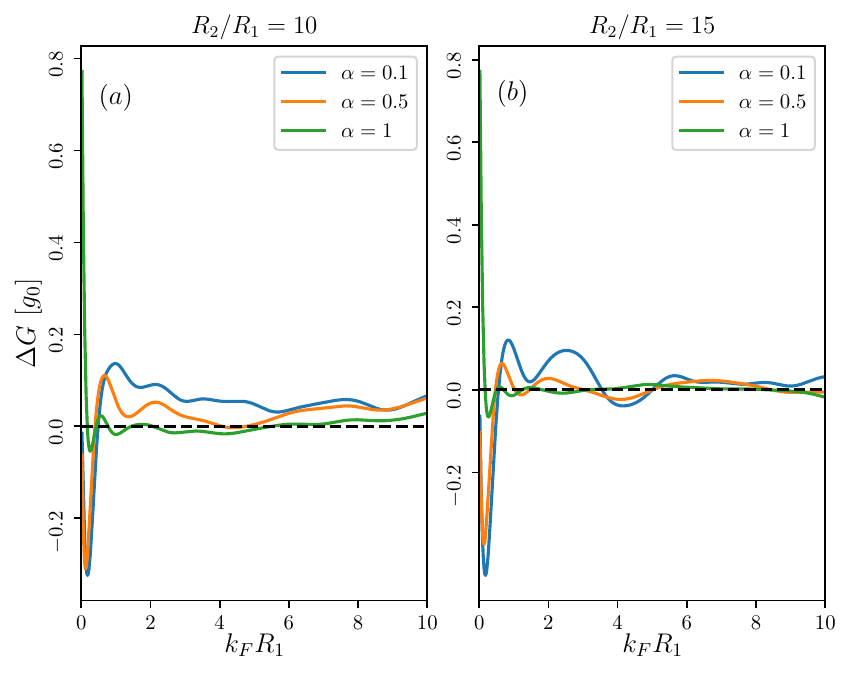}
\caption{(Color online) Magnitude of the conduction oscillations, $\Delta G = G(\Phi_0/2) - G(0)$, depicted as a function of the doping for selected values of $\alpha$ (specified for each line). Black dashed line shows the actual $\Delta G =0$. $(a)$ for $R_2/R_1=10$ and $(b)$ for $R_2/R_1=15$.}
\label{fig:DeltaG}
\end{figure}

We study the magnitude of conductance oscillations $\Delta G$. It is defined as the difference between $G(\Phi_0/2)$ and $G(0)$ as [\onlinecite{Ryc,Bou}]
\begin{eqnarray}
\Delta G = G(\Phi_0/2) - G(0).
\label{DelG}
\end{eqnarray}
Fig. \ref{fig:DeltaG} shows the magnitude of the conductance oscillations $\Delta G$ as a function of the doping $k_FR_1$ for two different radii ratios, namely $R_2/R_1 =10$ (see Fig. \ref{fig:DeltaG}$(a)$) and $R_2/R_1 =15$ (see Fig. \ref{fig:DeltaG}$(b)$). The absolute values of $\Delta G$ are relatively large not only in small vicinity of the charge neutrality point (zero doping), but also at larger doping. This means that for low Fermi wave vectors, the conductance oscillations are strong and more sensitive irrespective the value of $\alpha$, signaling the importance of transport via the evanescent waves. This feature has a similarity to that in graphene [\onlinecite{Ryc,Bou}]. Notice that $\Delta G$ (Eq. (\ref{DelG})), is governed by only a few modes for which $k_FR_1 \approx \nu(j)$ (see Eq. (\ref{nu})) and thus $T_j$'s are neither $\approx 0$ nor $\approx 1$. A systematic growth of $\Delta G$ with the ratio $R_2/R_1$ is visible for $k_F \to 0$ for $\alpha < 1$ cases, which is consistent with the corresponding results for graphene as shown in Refs. [\onlinecite{Ryc2,Ryc3,Ryc,Bou,Kat}]. $\Delta G$ shows oscillates around the zero value (black dashed line). Positive or negative values indicate increase or decrease in the conductance relative to the flux-free state. As $k_FR_1$ increases, the oscillations gradually dampen, becoming smaller and more stable at large $k_FR_1$ for all values of $\alpha$. This suggests that at higher values of Fermi energy, the quantum interference effects diminish. Further, the geometry (the ratio of radii) has a strong influence on the oscillation magnitudes, with larger ratios damping the oscillations more quickly. This implies that changing the geometry (tuning the disk region) of the system alters the sensitivity of the conductance to the external magnetic flux, doping, and $\alpha$.

For each values of the $R_2/R_1$ ratio, there is a unique series of discrete doping values where $\Delta G$ crosses the zero line, resulting in $G(\Phi_i)$ being constant (that is equal to $G[0]$). Moreover, for a given $R_2/R_1$, increasing $\alpha$ leads to an increase in the number of nodes. As shown in Fig. \ref{fig:DeltaG}$(a)$, for $R_2/R_1 = 10$, $\alpha = 0.1$ has only one node, while $\alpha = 0.5$ and 1 exhibit 3 and 6 nodes, respectively. Similarly, for $R_2/R_1 = 15$, the number of nodes increases to 3, 6, and 7 corresponding to $\alpha=0.1$, 0.5, and 1, respectively (see Fig. \ref{fig:DeltaG}$(b)$). For example, when $R_2/R_1 = 10$ and 15, the nodes of $\Delta G$ correspond to certain specific values of $(k_FR_1)_{\Delta G = 0}$ are listed in Table I.

\begin{table}[h!]
\begin{center}
%\label{Table1}
\begin{tabular}{|c|c|c|c|c|c|c|c|}
\hline
\multicolumn{8}{|c|}{$R_2/R_1 = 10$} \\
\hline
$\alpha$ & 1st & 2nd & 3rd & 4th & 5th & 6th & 7th\\
\hline
0.1 & 0.49 & - & - & - & - & - & - \\
\hline
0.5 & 0.40 & 4.09 & 4.82 & - & - & - & -\\
\hline
1 & 0.172 & 0.43 & 0.775 & 1.455 & 2.074 & 5.534 & -\\
\hline
\multicolumn{8}{|c|}{$R_2/R_1 = 15$} \\
\hline
$\alpha$ & 1st & 2nd & 3rd & 4th & 5th & 6th & 7th\\
\hline
0.1 & 0.56 & 3.55 & 5.00 & - & - & - & -\\
\hline
0.5 & 0.467 & 1.028 & 1.46 & 3.014 & 5.00 & 8.445 & -\\
\hline
1 & 0.174 & 0.504 & 0.701 & 1.251 & 1.796 & 3.194 & 8.149\\
\hline
\end{tabular}
\caption{Values of $k_FR_1$ for which $\Delta G = 0$ for $R_2/R_1 = 10$ and $R_2/R_1 = 15$.}
\end{center}
\end{table}

Prior to the first node, for $\alpha < 1$, we have $\Delta G < 0$, however, for $\alpha =1$, $\Delta G > 0$. Beyond that, the sign of $\Delta G$ alternates with growing values of $k_FR_1$ for each node, as indicated in Figs. \ref{fig:DeltaG}$(a)$ and $(b)$. It is also visible in Fig. \ref{fig:DeltaG} that the occurrence of the nodal points is irregular, as one can expect since $\Delta G$ can be regarded as originating from the behaviour of the Bessel function. For example, the typical separation between the first node of $\Delta G$ for $\alpha =0.1$ (see Table I) can be approximated as $\Delta k_FR_1 \approx 0.5$, which corresponds to $R_1 \approx 80$ nm, (nearly consistent with the results of graphene [\onlinecite{Ryc}]), to the energy interval of $\Delta E_f/k_B \approx 50$K (with the Boltzmann constant $k_B$). In turn, the conductance oscillations should be observable in comparable or higher temperatures then the standard AB effect in graphene rings [\onlinecite{Rus, Stam}].

\begin{figure}[h!]
\centering
\includegraphics[width=\linewidth]{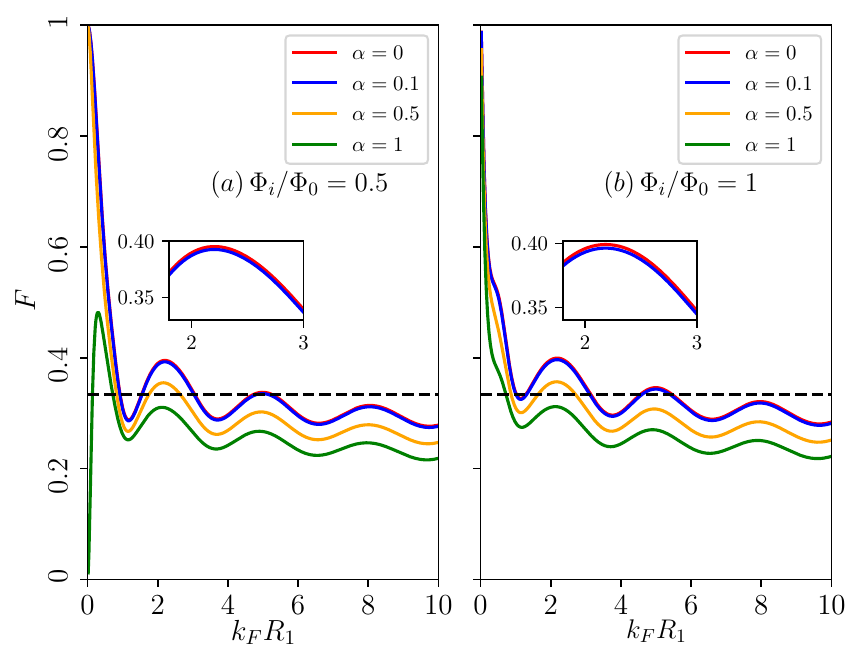}
\caption{(Color online) Fano factor $F$ as a function of the doping $k_FR_1$ for the radii ratio $R_2/R_1 = 10$ and $(a)$ for $\Phi_i/\Phi_0=0.5$ and $(b)$ for $\Phi_i/\Phi_0 = 1$ ($\Phi_i$ being the threaded flux). Different lines corresponding to $\alpha=0$ (red line), $\alpha=0.1$ (blue line), $\alpha=0.5$ (orange line), and $\alpha=1$ (green line). The black dashed line denotes diffusive transport regime. In the inset we show the zoomed version.}
\label{fig:Fano}
\end{figure}

\section{Fano factor}
\label{Fano1}
In general, the electrical current through nanostructures exhibits time-dependent fluctuations due to the discrete nature of electric charge [\onlinecite{Bla}]. These fluctuations, known as the shot noise, and the ratio of the shot noise power to the average electrical current is known as the Fano factor [\onlinecite{ker}]. The Fano factor assumes universal values depending on the electronic transport regime in tunnel junctions, such as, it corresponds to a Poissonian noise with $F = 1$ [\onlinecite{Ede}], for diffusive transport, it is $F = 1/3$ [\onlinecite{Nag,Been}], and in ballistic transport, $F = 0$ [\onlinecite{Bla}]. With the rise of graphene electronics, it was found that the conductivity of a graphene strip reaches its minimum at zero bias condition, causing the shot noise power to peak. Further, Ref. [\onlinecite{Two}] demonstrated that the Fano factor for a graphene strip is $1/3$, matching the value for a diffusive transport. This result stems from the non-classical transport in graphene, governed by the Dirac equation, and has been experimentally verified in Refs. [\onlinecite{Dan,Dic}]. Further, it was conceived that the energy of the Fano factor peaks track the locations of the Dirac points in the Brillouin zone [\onlinecite{Lima}].

In our case, the dominant contribution to the conductance fluctuation is due to the external magnetic flux. It would be interesting to explore how the conductance of an $\alpha$-$T_3$ disk responds to the external flux by examining the Fano factor, which defined as the following ratio,
\begin{eqnarray}\label{Fano}
F = \frac{\sum_j[T_{\nu(j)}(1-T_{\nu(j)})]}{\sum_j T_{\nu(j)}}.
\end{eqnarray}
where $\sum_j$ denotes the sum over all conducting modes and $T_{\nu(j)}$ is the transmission probability as defined in Eq. (\ref{Trans}) The results obtained so far can be numerically analyzed under suitable choices for the physical parameters. The results are shown in Fig. \ref{fig:Fano} with two different values of the flux, namely, $\Phi_i/\Phi_0 = 0.5$ (see Fig. \ref{fig:Fano}$(a)$) and 1 (see Fig. \ref{fig:Fano}$(b)$). Similar to the strip geometry [\onlinecite{Two}], the conductance minimum at $k_FR_1 \to 0$ corresponds to a maximum (for $\alpha < 1$) or a minimum (for $\alpha = 1$) value for the Fano factor, $F$. Further, $F$ acquires high values near $k_FR_1 \to 0$, which decreases sharply followed a series of oscillations as $k_FR_1$ increases. The amplitudes of the oscillations diminishes as $\alpha$ increases, and the curves for higher $\alpha$ (towards the dice limit) tend to flatten more. The black dashed line corresponds to $F=1/3$, which denotes a value, where the transport is fully diffusive. Our results shows that for all $\alpha$ and higher $k_FR_1$, we obtain the Fano factor assuming values closer to the diffusive limit. However, for the small $k_FR_1$ and $\alpha <1$, we get $F \sim 1$ that underscores a Poissonian transport regime [\onlinecite{Ede}] dominated by independent tunneling events, with either little or no correlation between the electron transmissions. The above result ($F\sim 1$) contrasts with the regime where $F<1$, which denotes diffusive (or sub-Poissonian) transport, where the electron correlations or scattering play a more significant role. Small values of $k_FR_1$ mean that the size of the system is smaller compared to the Fermi wavelength, thereby leading to a regime where the quantum effects dominate, and the transport is primarily dominated by random tunneling events.

Further, different flux ratios threaded, for example $\Phi_i/\Phi_0 = 0.5$ with $\alpha <1$, the system is likely to be weakly coupled, which contributes to the random (uncorrelated) nature of the transport. However, this is not the case for $\alpha=1$, where the Fano factor changes from $F \sim 0$ to nonzero values with increasing $k_FR_1$. This indicates a departure from the ballistic regime to a pseudo-diffusive transport ($F \to 1/3$) for large values of $k_FR_1$ as shown in Fig. \ref{fig:Fano}$(a)$. Moreover, for $\Phi_i/\Phi_0 = 1$, we observe a transition from Poissonian transport ($F \sim 1$) to a pseudo-diffusive regime with increasing doping, for all values of $\alpha$, as shown in Fig. \ref{fig:Fano}$(b)$. These results differ from those in graphene, where transport is ballistic in high-purity, mesoscopic samples [\onlinecite{Novo, Mia}] and diffusive or pseudo-diffusive when disorder and impurities come into play [\onlinecite{Two, Du,Mor,Muc}]. Here, by tuning the doping level and the external flux, we can modulate the transport properties and the correlations between electrons involved in the transport process.

The magnetic flux alters the phase of the electronic wavefunctions via the AB effect. At $\Phi_i/\Phi_0 = 0.5$, phase interference can either enhance or suppress the transport channels, depending on the value of $\alpha$. For $\alpha = 1$ and low doping, transport suppression is significant, resulting in nearly a noiseless transport ($F \sim 0$). In contrast, other values of $\alpha$ lead to a larger noise due to the following reasons. This happens because, when $\alpha = 1$, the path lengths between the $A$ to $B$ and $A$ to $C$ sublattices are equal, causing the flat band to become localized and preventing it from carrying current. As a result, the transport is dominated by the Dirac cones, which exhibit ballistic behavior, similar to that of graphene. When $\alpha < 1$, the differing path lengths between the $A$ to $B$ and $A$ to $C$ sublattices diminish flat-band the localization, causing higher fluctuations in the current due to less coherent transport. This leads to higher noise levels and the Fano factor approaches the Poissonian limit.

\section{Comparison between Graphene and $\alpha$-$T_3$ Corbino disks}
\label{Compare}
While studies on graphene Corbino disks exist in literature, its generalization in terms of an $\alpha$-$T_3$ structure has so far been ignored. In this section, we compare and contrast the properties of a graphene and the $\alpha$-$T_3$ Corbino disks, thereby emphasizing the highlights of our work. In a particular setup, the magnetoconductance studies of a graphene Corbino disk in presence of a uniform magnetic field reveals conductance oscillations with the flux threading the disk area (denoted by $\Phi_d$) with a period of $\Phi_0 = \frac{2h}{e} ln(\frac{R_{out}}{R_{in}})$ (where $R_{out}$ and $R_{in}$ are the outer and inner radii of the disk), the result clearly differs from the usual oscillations with one flux quantum. Additionally, at low but finite doping, these oscillations persist upto a limit of $|\Phi_d|\leq \Phi_d^{\rm{max}}$, beyond which the oscillatory behaviour vanishes and the conductance gets suppressed. In addition to observation of diffusive transport, ballistic transport has also been observed in this graphene Corbino disk configuration [\onlinecite{Ryc3, Kat}]. In an alternate scenario, when the disk subjected to a solenoidal magnetic potential (similar to our setup), the graphene Corbino disk configuration exhibits smooth Aharonov-Bohm type conductance oscillations in the ballistic transport regime, and the conductance displays step-like features with increasing doping [\onlinecite{Ryc, Bou1, Bou, Kama, Rut}].

Moving to the $\alpha$-$T_3$ Corbino disk, where an additional atom is located at the center of each hexagon connected via tunable hopping with one of the neighbours possesses a zero-energy flat band that may play a decisive role in shaping the transport characteristics. Enroute to the perfect dice limit ($\alpha=1$), in contrast to graphene ($\alpha=0$), the step-like conductance feature diminishes as $\alpha$ increases. The conductance of our $\alpha=0$ case of the $\alpha$-$T_3$ lattice is distinct from the graphene of the ref. [\onlinecite{Ryc}] as shown in Fig. \ref{fig:GvskF}$(b)$. This distinction arises due to the \(3 \times 3\) structure of the \(\alpha\)-\(T_3\) Hamiltonian, which hosts a zero-energy flat band that persists even at \(\alpha=0\), in sharp contrast to the two-band Dirac model for graphene. The presence of the flat band fundamentally differentiates the \(\alpha=0\) case of the \(\alpha\)-\(T_3\) lattice from graphene. As evident from the DOS, even if $\alpha=0$ case of $\alpha$-$T_3$ lattice, there are available states present at the zero energy (corresponding to the flat band). The flat band begins to influence both the DOS and eventually that would enhance the transport properties. An important difference being the DOS becomes more concentrated at zero energy due to the flat band, and features such as the step-like feature in the conductance spectrum, which was prominent in graphene progressively smoothens out with $\alpha$. This can lead to destructive interference along certain paths. Furthermore, the flat band alters phase coherence and scattering pathways, ultimately modifying the wavefunctions and affecting the quantum phase of carriers traveling in the disk. Again, the lattice structure of the $\alpha$-$T_3$ lattice consists of three distinct sublattices: $A$, $B$, and $C$. These sublattices are connected by two different hopping amplitudes, $t$ (between the $A$ and $B$ sublattices) and $\alpha t$ (between the $B$ and $C$ sublattices). During electron transmission, an electron can propagate between two $B$ sublattice sites via two distinct paths, one is characterized by hopping strength, $t$, while the other by $\alpha t$. Whereas, for the $A$ and $C$ sublattices, the available hopping paths are limited to $t$ for the $A$ sublattice and $\alpha t$ for the $C$ sublattice, respectively. This path difference induces a phase difference, giving rise to interference effects for any nonzero value of $\alpha$. This interference mechanism due to anisotropic hopping is a fundamental distinction from the two band graphene case. Interestingly, in the dice lattice limit ($\alpha = 1$), although the flat band is there, the identical hopping between the $B\to C$ and $A\to B$ reduces the contrast between the hopping pathways, leading to a conductance behavior that in some way mimics that of graphene (see Fig. \ref{fig:GvskF}$(c)$). However, the conductance features retain the unique signatures of the flat band in the quantum interference phenomena. Further, the transmission characteristics and the conductance behavior vary with \(\alpha\), with the conductance increasing as \(\alpha\) increases. In certain parameter regime, namely, the \textit{`thick'} disk limit (large $R_2/R_1$) and small values of $\alpha$, higher harmonic contributions emerge in the conductance profile. The physical explanations of which are provided in the preceding section. It turns out that the presence of the zero-energy flat band significantly alters the transport characteristics compared to graphene. This result may be counter intuitive as the carriers pertaining to the flat band have zero velocity and should have negligible role in transport. Finally, a statistical analysis yields the Fano factor to vary with both $\alpha$ and the magnetic flux, indicating emergence of distinct transport characteristics. It may be seen that graphene and the dice lattice demonstrate similar behaviour, however any intermediate $\alpha$ show distinct features from the limiting cases due to the aforementioned path differences in an $\alpha$-$T_3$ structure. It is relevant to mention that the Fano factor for our $\alpha=0$ case of the $\alpha$-$T_3$ lattice is different from the Fano factor of the ref. [\onlinecite{Bou1}] and is shown via the red curves in Fig. \ref{fig:Fano}. These differences open up possibilities for fabricating novel transport devices in flat-band systems, such as hydrodynamic electron magnetotransport and enhanced (and tunable) thermoelectric properties of the $\alpha$-$T_3$ Corbino disk.

\section{Conclusion}
\label{Con}
We have studied the $\alpha$-$T_3$ lattice in the form of a disk with inner radius $R_1$ and outer radius $R_2$, subjected to a magnetic flux $\Phi_i$ which perceived as an electron pump by Laughlin and should demonstrate quantum Hall effect at the low temperature. Utilizing the unique geometry of the Corbino disk, we have conducted analytical calculations leading to exact solution to the stationary Dirac equation. Also, we determined the transmission probability of an electron with a given value of angular momentum tunneling through the $\alpha$-$T_3$ Corbino disk. Further, we have computed a few relevant physical quantities, such as, the conductance $G$ and the Fano factor $F$.

Our numerical results were expressed in terms of the doping level $k_FR_1$, ratio of the outer to inner radii $R_2/R_1$, magnetic flux $\Phi_i$, and $\alpha$. The conductance as a function of the magnetic flux threading the disk shows periodic oscillation of the AB kind. However, the oscillation patterns and the amplitudes hugely depend upon the system parameters ($\alpha$, $k_FR_1$ etc.). Additionally, we observe higher harmonic modes at lower $\alpha$ values and larger radii ratio. 

Notably, away from the charge-neutrality point ($k_FR_1=0$), the conductance oscillations exhibit a substantial magnitude. Regardless of the value of $\alpha$, these oscillations are strong and highly sensitive, highlighting the role of the evanescent waves in the mechanism leading to electron transport. Moreover, the conductance oscillates around zero, reflecting the influence of the geometry, magnetic flux, doping, and $\alpha$. We have also computed the Fano factor, which reveals distinct behavior for $\alpha < 1$ ($\alpha$-$T_3$) and $\alpha = 1$ (dice). For $\alpha < 1$, the transport shifts from Poissonian to pseudo-diffusive as the doping increases. In contrast, for $\alpha = 1$, the Fano factor transitions from the ballistic regime to pseudo-diffusive transport at large $k_FR_1$.

\section*{ACKNOWLEDGMENTS}
M. I. sincerely acknowledges the financial support from the Council of Scientific and Industrial Research (CSIR), Govt. of India (File No. 09/731(0172)/2019-EMR-I). The authors thank the anonymous referee for useful comments.


\begin{thebibliography}{55}
\bibitem{Bipar}
Even with the central atom present, the $\alpha$-$T_3$ lattice is still bipartite. As shown in Fig. \ref{fig:sys}, $A$, $B$, and $C$ atoms represent the three basis sites of a unit cell. The coordination number is 3 for the $B$ and $C$ sublattices, while it is 6 for the $A$ sublattice. The coordination number of one subset of sites is distinct than the other. We refer to this as an asymmetry.
\end{thebibliography}
\end{document}